\documentclass[11pt]{article}
\usepackage{amsfonts}
\usepackage{amssymb}
\usepackage{amsmath,amssymb}
\usepackage{bbm}
\usepackage{tikz}

\def\slasha#1{\setbox0=\hbox{$#1$}#1\hskip-\wd0\hbox to\wd0{\hss\sl/\/\hss}}

\def\periodb#1{\setbox0=\hbox{$#1$}#1\hskip-\wd0\hbox to\wd0{-}}

\def\sfrac#1#2{{\textstyle\frac{#1}{#2}}}
\newcommand{\unit}{\mathbbm{1}}   

\newcommand{\CA}{\mathcal{A}}    
\newcommand{\CJ}{\mathcal{J}} 
\newcommand{\CT}{\mathcal{T}} 
\newcommand{\CF}{\mathcal{F}}   
\newcommand{\CG}{\mathcal{G}}  
\newcommand{\CE}{\mathcal{E}}
\newcommand{\CH}{\mathcal{H}}
\newcommand{\CR}{\mathcal{R}}
\newcommand{\Z}{\mathbb{Z}}   
\newcommand{\R}{\mathbb{R}}     
\newcommand{\C}{\mathbb{C}}     
     
\newcommand{\im}{\mathrm{i}} 
\newcommand{\Id}{\mathrm{Id}} 
\newcommand{\dd}{\mathrm{d}}     
\newcommand{\dpar}{\partial}     

\newcommand{\+}{\dagger}

\newcommand{\sU}{\mathrm{U}}

\newcommand{\sSO}{\mathrm{SO}}     
\newcommand{\sGL}{\mathrm{GL}}  
 
\newcommand{\rso}{\mathrm{so}} 
\newcommand{\ru}{\mathrm{u}}     

\newcommand{\al}{{{\alpha}}} 

\newcommand{\vr}{{{\varrho}}}
\newcommand{\vph}{{{\varphi}}} 

\newcommand{\xh}{\hat{x}} 
\newcommand{\ph}{\hat{p}}
\newcommand{\fh}{{\hat{f}}}

\newcommand{\zb}{{\bar{z}}}

\newcommand{\Av}{{A_{\sf{vac}}}}
\newcommand{\Fv}{{F_{\sf{vac}}}}

\newcommand{\qv}{{q_{\sf{v}}}}
\newcommand{\ql}{q_l^{}}

\newcommand{\und}{\quad\mbox{and}\quad}
\newcommand{\with}{\quad\mbox{with}\quad}

\makeatletter

\@addtoreset{equation}{section}
\makeatother

\begin{document}
\begin{titlepage}
\setcounter{page}{0}
.
\vskip 15mm
\begin{center}
{\LARGE \bf Antiparticles in non-relativistic quantum mechanics}\\
\vskip 2cm
{\Large Alexander D. Popov}
\vskip 1cm
{\em Institut f\"{u}r Theoretische Physik,
Leibniz Universit\"{a}t Hannover\\
Appelstra{\ss}e 2, 30167 Hannover, Germany}\\
{Email: alexander.popov@itp.uni-hannover.de}
\vskip 1.5cm
\end{center}
\begin{center}
{\bf Abstract}
\end{center}

\vskip 0.5cm

Non-relativistic quantum mechanics was originally formulated to describe particles. Using ideas from the geometric quantization approach, we show how the concept of antiparticles can and should be introduced in the non-relativistic case without appealing to quantum field theory. We discuss this in detail using the example of the one-dimensional harmonic oscillator.


\end{titlepage}


\section{Introduction and summary}

\noindent  {\bf Prequantum gauge theory.} Let us consider an even dimensional manifold $X$ with a symplectic two-form $\omega_X$ ($=\dd\theta_X$ locally) as the phase space of Hamiltonian mechanics. 
A non-relativistic classical particle is a point in space $X$ moving along a trajectory in $X$ given by a Hamiltonian vector field $V_H^{}$ generated by a function $H$ (Hamiltonian) with an evolution parameter $t$.
In the approach of geometric quantization (see e.g. \cite{Sni, Wood}), which is a neat mathematical reformulation of the canonical quantization procedure, the transition from $(X, \omega_X)$ to quantum mechanics is carried out through
\begin{itemize}
\item the introduction of the principal U(1)-bundle $P(X, \sU(1)_{\sf{v}})$ over $X$ and the associated complex line bundle $L_{\sf{v}}\to X$ with connection\footnote{We use the natural units with $\hbar =c=1$.}  $\Av =\im\,\theta_X$ and curvature $\Fv =\im\,\omega_X$,
\item the introduction of polarization $\CT$ on $X$, i.e. integrable Lagrangian subbundle of the complexified tangent bundle $T^\C X$ of $X$. 
\end{itemize}
Both $\Av$ and $\Fv$ take values in the Lie algebra $\ru (1)_{\sf v}=\rm{Lie}\sU(1)_{\sf v}$. The abbreviations ``$\sf v$" and ``$\sf vac$" here mean ``vacuum" since $\theta_X$ and $\omega_X$ have no sources and define a symplectic structure on $X$.

The polarization $\CT$ makes it possible to impose on sections $\psi$ of the bundle $L_\C^+:=L_{\sf{v}}$ the condition $\dd_\CT^{}\psi =0$  of independence from a ``half" of the coordinates on $X$.  The space $\CH$ of polarized sections of $L_\C^+$ is the quantum Hilbert space and functions $\psi\in\CH$ are the ``wavefunctions" of quantum mechanics. Accordingly, the bundle $L_\C^+$ with a polarization $\CT$ is a quantum bundle encoding the basic quantum mechanical (QM) information. By definition, a non-relativistic quantum particle is a polarized section $\psi_+\in \Gamma (L_\C^+, \CT)$ of the complex line bundle $L_\C^+$ over $X$.

\medskip

\noindent  {\bf QM as gauge theory.} The above two steps (bundle $L_\C^+$, polarization $\CT$) towards the introduction of quantum mechanics do not encounter obstacles, because it is simply the introduction of a bundle over $X$ with some conditions on $X$ and $L_\C^+$, for example, the holomorphicity condition. Moreover, it is not difficult to show that covariant derivatives in $L_\C^+$ can be chosen so that they coincide with the standard operators of coordinates and momenta. Having covariant derivative  $\nabla_A$ acting on polarized sections $\psi_+$ of the bundle $L_\C^+$, one can construct quantum Hamiltonians as is customary in differential geometry. However, geometric quantization, like any other quantization scheme, also attempts to define a map
\begin{itemize}
\item functions $f\in C^\infty (X)\ \to\ $ operators $\fh$ on a Hilbert space $\CH$
\end{itemize}
which should satisfy some axioms \cite{Sni, Wood}. But this does not work because of the problem with operator ordering, inconsistency with polarization and many other problems. For this reason we do not use either the map $f\mapsto\fh$ or the metaplectic correction, but keep only first two steps and consider quantum mechanics as the theory of fields $\psi_+\in \Gamma(L_\C^+, \CT)$, which are acted upon by covariant derivatives $\nabla_A$ with canonical connection $\Av = \im\,\theta_X$.

\medskip

\noindent  {\bf Antiparticles and quantum charges.} 
The logic of gauge theories tells us that if there is a complex line bundle $L_\C^+$, then we should also consider the complex conjugate bundle $L_\C^-:=\overline{L_\C^+}$ and associate the charges $\qv =1$ and $\qv =-1$ with the structure group U(1)$_{\sf v}$ of these bundles, which leads to the interpretation of sections of these bundles as particles and antiparticles. Observables are introduced not through the mapping $f\to\fh$, but using a metric on $X$ and covariant derivatives $\nabla_A$, which automatically gives the Weyl ordering.

Connection $A^{}_{\sf vac}=-\im \theta_X$ and curvature $F^{}_{\sf vac}=-\im \omega_X$ of the bundle $L_\C^-$ have opposite signs compared to those in the bundle $L_\C^+$. A non-relativistic quantum antiparticle is a polarized section 
$\psi_-\in\Gamma (L_\C^-, \CT^*)$ of the complex line bundle $L_\C^-$. In Hamiltonian mechanics, it corresponds to a point in the symplectic manifold $(X, -\omega_X)$ with the evolution parameter $-t$. 

  The fibres of the principal bundle $P(X, \sU(1)_{\sf v})$ are circles $S^1$ parametrized by elements $e^{\im\theta}\in\sU(1)_{\sf v}\subset\C$, where $\C$ is a typical fibre of the associated  complex line bundle $L_\C^+$ with a coordinate $\psi_+=\rho_+e^{\im\theta}\in\C$ on the fibre. A circle $S^1$ is a closed curve on $\C$ around the point $\psi_+=0$ and it has the winding number $\qv=1$. The complex conjugate bundle  $L_\C^-$ has on its fibres a complex coordinate $\psi_-=\rho_-e^{-\im\theta}$ and the winding number $\qv =-1$. 

In general, the winding number of a curve in $\C$ is the total number of times that curve travels counterclockwise around a fixed point for $\qv=r$ and clockwise for $\qv =-s$, $r,s=0,1,...$. For $S^1\subset\C$, these curves correspond to one-dimensional representations of the group U(1) and $\C^*$, defined by the homomorphisms
\begin{equation}\label{1.1}
f_\qv :\ S^1\to S^1\with f_\qv(e^{\im\theta})=e^{\im\qv\theta}\quad \mbox{for}\ \ \qv\in\pi_1(S^1)=\Z\ ,
\end{equation}
and, respectively,
\begin{equation}\label{1.2}
\psi_+\to\psi_+^r \quad \mbox{for}\ \ \qv=r\ge 0\und \psi_-\to\psi_-^s \quad \mbox{for}\ \qv=-s\le 0\ .
\end{equation}
The first mapping in \eqref{1.2} defines the tensor product of bundles $(L_\C^+)^{\otimes r}$ with $\qv=r$, and the second mapping in  \eqref{1.2} defines the tensor product of  bundles $(L_\C^-)^{\otimes s}$ with $\qv =-s$.  We will call $\qv$ a {\it quantum charge}, and sections of bundles with $\qv\ge 0$ are {\it particles}, and those with $\qv\le 0$ are {\it antiparticles}. Sections with $\qv= 0$ are  neutral particles, they are sections of the diagonal subbundle in the bundle $L_\C^+\oplus L_\C^-$.
We emphasize that all quantum particles have a quantum charge, including those that have zero electric charge.

\medskip

\noindent  {\bf Problem statement.} In classical Hamiltonian mechanics and the corresponding nonrelativistic quantum mechanics (QM) there is no concept of antiparticles. These theories describe only particles, and the number of particles $\qv =r$ is given as initial data and cannot change.\footnote{Particles can be created and destroyed only in quantum field theory.} In Newtonian mechanics, non-relativistic particles are described by a phase space $X=T^*\R^3 =\R^3\times\R^3$ parametrized by coordinates $x\in \R^3$ and momenta $p\in\R^3$. Antiparticles are introduced after the transition to relativistic QM with phase space $T^*\R^{3,1} =\R^{3,1}\times\R^{3,1}$, where positive frequency solutions of free wave equations are associated with particles, and negative frequency solutions are associated with antiparticles. This interpretation is carried over to quantum field theory (QFT), where the coefficients in the expansion in positive and negative frequency basis solutions are replaced by operators.

In the non-relativistic limit $c\to\infty$, positive-frequency solutions $\psi_+$ of relativistic equations are identified with the wave functions of free non-relativistic particles with energy $E=p^2/2m$, and the analogous limit of negative-frequency solutions is declared non-physical. However, antiparticles do exist at low velocities $v<<c$. That is why attempts have been made repeatedly to find a transition from QFT to non-relativistic QM while preserving the concept of antiparticles (see e.g. \cite{Padm, Sa} and references therein), but all these attempts were not considered successful.

Returning to formulae  \eqref{1.1} and  \eqref{1.2}, we note that differential geometry asserts that if a particle is associated with sections $\psi_+$ of the bundle $L_\C^+$, then sections of the dual bundle must be associated with antiparticles (opposite charges $\qv =+1$ and $\qv=-1$). It is well known that if $\CE$ is a rank $n=1,2,...$ complex vector bundle then there are three more complex vector bundles: the complex conjugate bundle $\bar\CE$, the dual bundle $\CE^{\sf{v}}$ and the dual of the complex conjugate bundle $\bar\CE^{\sf{v}}$.
For example, if quarks are described by a complex vector bundle $\CE$ of rank 3 as $\C^3$-columns, then antiquarks are described by the dual bundle  $\CE^{\sf{v}}$ as  $\C^3$-rows. If $\CE$ is a Hermitian complex vector bundle then bundles $\bar\CE$ and $\CE^{\sf{v}}$ are isomorphic as well as bundles $\CE$ and $\bar\CE^{\sf{v}}$. Bundles $\CE$ and $\bar\CE$ are not isomorphic and in particular $k$-th Chern class of $\bar\CE$ is given by $c_k(\bar\CE)=(-1)^k c_k(\CE)$ so that $c_1(\bar\CE)=- c_1(\CE)$.

The bundle $L_\C^+$ introduced in non-relativistic QM is a Hermitian complex line bundle \cite{Sni, Wood}. Therefore, we can and must introduce the complex conjugate bundle $L_\C^-$ and its sections $\psi_-$ will describe antiparticles when $\psi_-\ne\psi_+^*$. In this case, the energy $E$ of particles and antiparticles is always the same, since in the Schr\"odinger equations for dual objects $\im\dpar_t$ is replaced by  $-\im\dpar_t$. Differential geometry requires distinguishing particles and antiparticles by the quantum charge $\qv =\pm 1$ introduced above, and not by the sign of energy, and the definition in terms of bundles $L_\C^\pm$ is preserved for particles in any external fields and in the relativistic case. In particular, the energy of a free non-relativistic electron is positive, but for an electron in the field of central forces generated by a proton it is negative. Moreover, if the hydrogen atom is described by the function $\psi_+\in\Gamma (L_\C^+)$, then the positron in the field of forces generated by the antiproton is described by the function $\psi_-\in\Gamma (L_\C^-)$, and the experimentally observed energy levels of hydrogen and antihydrogen atoms are the same.

Using the differential geometric approach leads to the assertion that the density of conserved charges for positive and negative frequency complex solutions of the Klein-Gordon equation,
\begin{equation}\label{1.3}
\vr_\pm^{\sf KG}=\frac{\im\hbar}{2mc^2}\,\Bigl(\psi_\pm^*\dot\psi_\pm^{}-\dot\psi_\pm^*\psi_\pm^{}\Bigr)
\ \stackrel{c\to\infty}{\longrightarrow}\ \pm\psi_\pm^*\psi_\pm^{}\ ,\quad \dot\psi:=\frac{\dpar\psi}{\dpar t}\ ,
\end{equation}
is the density $\vr_\pm^{\sf Sch}=\pm\psi_\pm^*\psi_\pm^{}$ of charges $\qv=\pm 1$ associated with the bundles $L_\C^\pm$ and the continuity equations for the conjugated Schr\"odinger equations. And the interpretation of functions $|\psi_\pm^{}|^2=\psi_\pm^*\psi_\pm^{}$ as probability densities is secondary. This can be illustrated by the example of a one-dimensional harmonic oscillator, considered in this paper. Namely, the function $\psi_+^*\psi_+^{}$ is the quantum charge density in both the coordinate and holomorphic Segal-Bargmann representations. However, in the holomorphic representation, the function $\psi_+^*\psi_+^{}$ cannot be interpreted as a probability density (there is no $\sigma$-additivity).

The inadequacy of introducing the concept of a particle and antiparticle through the sign of the eigenvalues of the operator $\im\dpar_t$ is obvious when considering quantum fields in curved space-time, which is discussed in detail in \cite{BD}.  In general, Killing vectors, with the help of which one could define positive frequency solutions, do not exist at all, and this leads to the vagueness of the concept of a particle \cite{BD}. The definition in terms of bundles $L_\C^+$ and $L_\C^-$ removes this problem since being a particle or antiparticle means having a conserved quantum charge $\qv =+1$ or $\qv=-1$. Even in  Minkowski space, the definition via the operator  $\im\dpar_t$ does not work for non-free particle, for example, for a relativistic oscillator. Its covariant phase space is a homogeneous K\"ahler space PU(3,1)/U(3) which is not fibred over either coordinate space or the momentum space, and the solutions do not contain terms with $\exp(\pm\im Et)$. Instead, the oscillating particles and antiparticles are sections of the holomorphic $L_\C^+$ and, respectively, antiholomorphic $L_\C^-$ bundles over the manifold PU(3,1)/U(3) \cite{Popov}.

The differential-geometric view of quantum mechanics requires another conceptual change. The ``wavefunctions" $\psi_\pm^{}$ of particles and antiparticles are not functions, but sections of {\it vector} bundles $L_\C^\pm$. Therefore, they should be added not as scalars, but as vectors:
\begin{equation}\label{1.4}
\Psi =\psi_+^{}v_+^{}+\psi_-^{}v_-^{}\in L_\C^+\oplus L_\C^-\ ,
\end{equation}
where $v_\pm^{}$ are bases in the fibres of vector bundles $L_\C^\pm$ of rank one. Also in relativistic equations it is necessary to separate fields with $\qv =1$ and $\qv =-1$ as in \eqref{1.4}, which allows to get rid of the problem of negative energies.

In this paper we consider non-relativistic classical and quantum harmonic oscillator to illustrate all that has been said above. Our goal is not to say anything new about the oscillator described in textbooks. Using the example of an oscillator, we want to analyze what a particle and antiparticle with charge $\qv=\pm 1$ are at the classical and quantum levels. In addition, we focus on describing oscillators as charged particles in an Abelian gauge field with constant field strength $\Fv=\pm\im\omega_{\R^2}^{}$. In other words, we regard quantum mechanics as an Abelian gauge theory with a fixed background connection $\Av =\pm\im\theta_{\R^2}^{}$ with curvature $\Fv$ on bundles $L_\C^\pm$.

\section{Classical harmonic oscillator}

\noindent  {\bf Symplectic structure.} In classical mechanics, a simple harmonic oscillator is a particle of mass $m$ under the influence of a restoring (attractive) force $F=-m\omega^2x$ and the equation of motion
\begin{equation}\label{2.1}
\ddot x + \omega^2 x=0\ ,
\end{equation}
where $x\in\R$ is a coordinate, $\dot x:=\dd x/\dd t$, $\omega$ is the frequency and $t$ is the evolution parameter.

Let us consider the phase space $T^*\R =\R^2$ of oscillator with coordinate $x$ and momentum $p$ and define on it a symplectic 2-form
\begin{equation}\label{2.2}
\omega^{}_{\R^2}=\dd\theta^{}_{\R^2}=\dd p\wedge\dd x\quad \mbox{for}\ \ \theta^{}_{\R^2}=\sfrac12\,p\dd x -\sfrac12\,x\dd p
\end{equation}
with a bivector
\begin{equation}\label{2.3}
\omega^{-1}_{\R^2}=\dpar_x\wedge\dpar_p
\end{equation}
inverse to the 2-form  \eqref{2.2}. We define a function (Hamiltonian) on $\R^2$,
\begin{equation}\label{2.4}
H=\frac{p^2}{2m}+ \frac{m\omega^2x^2}{2} \ ,
\end{equation}
and associate with it the Hamiltonian vector field
\begin{equation}\label{2.5}
V_H=\omega^{-1}_{\R^2}(\dd H)=\dpar_pH\dpar_x - \dpar_xH\dpar_p = \frac{p}{m}\dpar_x - m\omega^2x\dpar_p\ .
\end{equation}
Then equation \eqref{2.1} can be rewritten in the Hamiltonian form
\begin{equation}\label{2.6}
\dot x=V_H(x)=\frac{p}{m}\und \dot p = V_H(p)=-m\omega^2x=F\ ,
\end{equation}
where $F$ is the restoring force.

\medskip

\noindent  {\bf Complex structure.} We introduce complex coordinates on $T^*\R=\R^2$,
\begin{equation}\label{2.7}
z=\frac{1}{\sqrt 2}\,(x-\im w^2p)\und\zb =\frac{1}{\sqrt 2}\,(x+\im w^2p)
\end{equation}
with derivatives
\begin{equation}\label{2.8}
\dpar_z=\frac{1}{\sqrt 2}\,\left(\dpar_x+\frac{\im}{ w^2}\dpar_p\right)\und\dpar_\zb =\frac{1}{\sqrt 2}\,\left(\dpar_x-\frac{\im}{ w^2}\dpar_p\right)\ .
\end{equation}
Here $w\in\R^+$ is a length parameter defined as\footnote{The dependence of all quantities on Plank's constant $\hbar$ is not important for us, so we use the natural units with $\hbar =c=1.$}
\begin{equation}\label{2.9}
w^2=\frac{1}{m\omega}
\end{equation}
so that $[w^2p]=[$length$]=[x]$.

Note that $\dpar_x$ and $\dpar_p$ form the basis of the tangent space  $V=\R^2$ to $T^*\R$ and on it we can introduce an endomorphism $J\in\mbox{End}(V)$ defined by formulae
\begin{equation}\label{2.10}
J(\dpar_z)=\im\dpar_z\und J(\dpar_\zb)=-\im\dpar_\zb\ .
\end{equation}
It is easy to see that
\begin{equation}\label{2.11}
J\left(\frac{\dpar}{\dpar x^a}\right)= J_a^b\,\frac{\dpar}{\dpar x^b}\with J^1_2=-1,\ J^2_1=1\ \Rightarrow\ J=\begin{pmatrix}0&-1\\1&0\end{pmatrix}\ ,
\end{equation}
where $x^1:=x$ and $x^2:=-w^2p$. For matrix $J\in\mbox{End}(V)$, condition $J^2=-\unit_2$ is satisfied and $J$ is called a complex structure on $\R^2$, so that $(\R^2, J)\cong\C$ is a complex space with coordinate $z$. From  \eqref{2.10} it also follows that $(\R^2, -J)=\bar\C$ with the complex conjugate coordinate $\zb$, i.e. complex conjugation is equivalent to replacing $J\to -J$ in \eqref{2.11}.

\medskip

\noindent  {\bf Harmonic oscillators.} In terms of the complex coordinate $z$ on $\R^2$, the equations \eqref{2.6} and their solutions have the form
\begin{equation}\label{2.12}
\dot z=\im\omega z\ \Rightarrow\ z=e^{\im\omega t}z_0\ ,
\end{equation}
where $z_0$ specifies the initial values of $x$ and $p$. Note that $z_0=0$ gives a trivial solution $z=0$ and therefore the space $\R^2\backslash\{0\}=\C\backslash\{0\}$ is usually considered as the phase space of the oscillator. We identify this solution with a particle, it has a quantum charge $\ql =1$ coinciding with the winding number of the circle $e^{\im\tau}$  with $\tau =\omega t$ in \eqref{2.12}. To indicate the sign of this winding number, we rewrite solution \eqref{2.12} as
\begin{equation}\label{2.13}
z_+=e^{\im\omega t}z_0\with z_0=\frac{1}{\sqrt 2}\,(x_0-\im w^2p_0)=\rho_0e^{\im \vph_0},\quad w^2p_0=\frac{v_0}{\omega}\ ,
\end{equation}
where $v_0$ is the initial velocity. 

Let us now make the substitution $t\to -t$ in \eqref{2.6}. This is equivalent to replacing $p\to -p$ and changing the sign of the complex structure on $\R^2$, 
\begin{equation}\label{2.14}
J_p^x=\frac{1}{w^2}\ \to\ J_p^x=-\frac{1}{w^2}\und J^p_x=-w^2\ \to\  J^p_x=w^2\ ,
\end{equation}
and therefore obtaining  a new complex coordinate
\begin{equation}\label{2.15}
z_-=\frac{1}{\sqrt 2}\,(x+\im w^2p)\ .
\end{equation}
In terms of $z_-$, equations \eqref{2.6} and their solutions have the form
\begin{equation}\label{2.16}
\dot z_-=-\im\omega z_-\quad\Rightarrow\quad z_-=e^{-\im\omega t}\zb_0^\prime\with \zb_0^\prime =\frac{1}{\sqrt 2}\,(x_0^\prime +\im w^2p_0^\prime)=\rho_0e^{-\im\vph_0^\prime}\ ,
\end{equation}
where $z_0^\prime$ does not necessarily coincide with $z_0$. This solution has $\ql =-1$, since it describes movement along $S^1$ clockwise. Thus, the reversal of time corresponds to a change $J\to -J$ in the sign of the complex structure, a change in the sign $\omega_{\R^2}\to -\omega_{\R^2}$ and a replacement of $\ql =1$ (particle) with $\ql =-1$ (antiparticle).

On the plane $(x^1, x^2)=(x, -w^2p)$ we have 
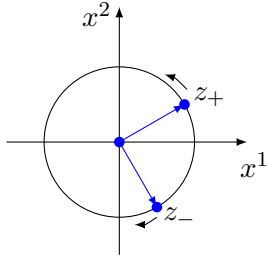
\begin{center}
\begin{tikzpicture}[>=latex]
\draw[->] (-1.5,0) -- (1.7,0); 
\draw[->] (0,-1.5) -- (0,1.8); 
\draw (0,0) circle (1cm); 
\fill[blue] (0,0) circle (2pt); 
\draw[blue, ->] (0,0) -- (0.866, 0.5); 
\draw[blue, ->] (0,0) -- (0.5, -0.866); 
\fill[blue] (0.866,0.5) circle (2pt); 
\fill[blue] (0.5, -0.866) circle (2pt); 
\node at  (1.2,0.6) {$z_+$};
\node at  (-0.3,1.7) {$x^2$};
\node at  (0.8,-1) {$z_-$};
\node at  (1.8,-0.3) {$x^1$};
\draw[->]  (0.9,0.7) .. controls (0.75,0.85) .. (0.6,0.9);
\draw[->]  (0.5,-1) .. controls (0.35,-1.1) .. (0.2,-1.1);
\end{tikzpicture}
\end{center}
and for $x, p$ we have
\begin{equation}\label{2.17}
\begin{array}{ll}
x(t)=x_0\cos\omega t +v_0\frac{\sin\omega t}{\omega}\ ,\quad p_0=mv_0\ ,\\[2pt]
p(t)=p_0\cos\omega t -m\omega x_0\sin\omega t\ ,
\end{array}
\end{equation}
where $x_0, p_0$ are initial data for $\ql =1$. For $\ql =-1$ we have \eqref{2.17} with $t\to -t$ and $(x_0, p_0)\to (x_0^\prime, p_0^\prime)$. Thus we get
\begin{equation}\label{2.18}
\begin{array}{ll}
\bigl(z_+=e^{\im\omega t} z_0 , \ \omega^{}_{\R^2},\ J,\ \ql=1\bigr)=\mbox{particle}\\[2pt]
\bigl(z_-=e^{-\im\omega t} \zb_0^\prime , \ -\omega^{}_{\R^2},\ -J,\ \ql=-1\bigr)=\mbox{antiparticle}
\end{array}
\end{equation}
Note that exactly the same difference as between $z_+$ and $z_-$ exists for vortices and antivortices on the plane $\R^2$; they are also associated with complex conjugate spaces and are characterized by integer winding numbers with $\ql >0$ and $\ql<0$ (see e.g. \cite{MS}).

In conclusion, we note that in \eqref{2.18} we consider a solution $z_+$ with positive and $z_-$ with negative frequency, as is customary in mathematics. In physics, it is often accepted the other way around, and to do this, you simply need to swap the complex and complex conjugate fields and solutions.

\medskip

\noindent  {\bf Symplectic reduction.} The complex structure \eqref{2.11} can be associated with the vector field
\begin{equation}\label{2.19}
\CJ = J_b^a x^b\dpar_a = x^1\dpar_2 - x^2\dpar_1= w^2p\dpar_x - \frac{x}{w^2}\,\dpar_p=\im (z\dpar_z - \zb \dpar_\zb ) =\dpar_\vph\ .
\end{equation}
This vector field defines the group U(1) of transformations of complex coordinates $z$ and $\zb$,
\begin{equation}\label{2.20}
\sU(1)\ni g=e^{\im\al\dpar_\vph}:\quad gz=e^{\im\al}z\und g\zb =e^{-\im\al}\zb\ ,
\end{equation}
corresponding to rotations on the plane $\R^2$ with winding numbers $\ql =1$ and $\ql =-1$, respectively. At the same time, comparing \eqref{2.5} and \eqref{2.19}, we see that
\begin{equation}\label{2.21}
V_H=\omega\CJ\ \Rightarrow\ \dot z_+=\omega\CJ (z_+)\und\dot z_-=\omega\CJ (z_-)\ ,
\end{equation}
i.e. the Hamiltonian of the oscillator has a geometric origin. This leads to a geometric interpretation of both equations in \eqref{2.21} and their solutions.

To see the geometry behind  \eqref{2.21}, we identify the dual space $\ru(1)^*$ of the Lie algebra $\ru(1)\cong\mbox{Lie}\,\sU(1)$ with the generator  \eqref{2.19} with the space $\R$. Then we can define the moment map
\begin{equation}\label{2.22}
\mu :\ \C\to \R\quad\mbox{as}\quad\mu (z)=z\zb =\rho^2_0=z_0\zb_0\in \R
\end{equation}
and the level surface
\begin{equation}\label{2.23}
S^1=\mu^{-1}(\rho_0)=\bigl\{z\zb =\rho_0^2\bigr\}
\end{equation}
is a circle. Group \eqref{2.20} preserves this circle and the Marsden-Weinstein symplectic reduction \cite{MW} of the space $\C\backslash\{0\}$ under the action \eqref{2.20} of the group U(1) is the quotient
\begin{equation}\label{2.24}
\C\backslash\{0\}//\sU(1) = \mu^{-1}(\rho_0)/\sU(1) =z_0\in\C\backslash\{0\}\ .
\end{equation}
Thus, equations \eqref{2.21} describe the reduction of the phase space $\C\backslash\{0\}$ of the oscillator to point $z_0$ (moduli space). We have the map
\begin{equation}\label{2.25}
\C\backslash\{0\}\to z_0\ ,
\end{equation}
and solution \eqref{2.13} specifies the circle \eqref{2.23} lying in the fibre of the projection \eqref{2.25}. The antioscillator is obtained for opposite signs of the symplectic and complex structures, as in \eqref{2.18}. Note that the energy of both solutions $z_\pm (t)$ is positive and equal to
\begin{equation}\label{2.26}
E=\frac{p^2}{2m}+\frac{m\omega^2x^2}{2}=\omega\gamma^2\quad\mbox{for}\quad\gamma^2=\frac{\rho^2_0}{w^2}\ ,
\end{equation}
where the real parameter $\gamma^2$ can take any non-negative value.

\medskip

\noindent  {\bf K\"ahler metric.} Having symplectic and complex structures on $\R^2$, we can introduce the K\"ahler metric on $\R^2$ by the formula
\begin{equation}\label{2.27}
g_{ab}=\omega_{ac}J_b^c\ \Rightarrow\ g=g_{11}(\dd x^1)^2 + g_{22}(\dd x^2)^2=\frac{1}{w^2}\,(\dd x^2 + w^4\dd p^2)\ .
\end{equation}
In what follows, we will use the rescaled metric
\begin{equation}\label{2.28}
\dd s^2_{\R^2}=w^2g = \dd x^2 + w^4\dd p^2 = 2\dd z\dd\zb\ ,
\end{equation}
which is not dimensionless, unlike \eqref{2.27}.

\section{Quantum bundles $L_\C^\pm$}

\noindent  {\bf Gauge theory $\Rightarrow$ QM.} Repeating what was said in the introduction, we emphasize that we are not engaged in quantization in the spirit of the Dirac program \cite{Dirac} and are not considering the mapping $f\mapsto\fh$ of functions $f$ on phase space into quantum operators $\fh$. Instead, we consider a gauge theory on phase space described by the set $(L_\C^+, \Av , \CT)$, where the connection $\Av$ defines the canonical commutation relations (CCR), and the polarization $\CT$ defines the Hilbert space on which the CCRs are irreducibly realized. Note that the connection $\Av$ on $L_\C^+$ is given up to an automorphism of the bundle $L_\C^+$ and by choosing different automorphisms we can obtain coordinate, momentum, holomorphic or antiholomorphic representations that are unitarily equivalent by virtue of the Stone-von Neumann theorem. We will show how all these representations are obtained from the choice of polarization $\CT$ and automorphism from the group Aut$(L_\C^+)$ for the case of harmonic oscillators. All observables are introduced only through the operators of covariant derivatives and a metric on phase space. The field $\Av\in \ru(1)_{\sf v}$ enters in these covariant derivatives and determines the interaction of particles with vacuum.

\medskip

\noindent  {\bf Principal bundle $P(\R^2, \sU(1)_{\sf v})$.} Let us consider a Newtonian particle of mass $m$ in one-dimensional space $\R$. On its phase space $T^*\R =\R^2$, the symplectic structure \eqref{2.2}, the complex structure \eqref{2.7}-\eqref{2.11} and the metric \eqref{2.28} are given. This particle is a harmonic oscillator if we choose the Hamiltonian in the form \eqref{2.4}. This Hamiltonian defines the vector field \eqref{2.5} which has the geometric meaning of the generator \eqref{2.19} of the group SO(2)$\cong$U(1) acting on $\R^2\cong\C$ by rotations \eqref{2.20}. A particle oscillating in space $\R$ corresponds to a point $z_+(t)$ on phase space $\R^2$ moving in a circle \eqref{2.23} with a winding number $\ql =1$.  Antiparticles are described by a trajectory $z_-(t)$ in $\R^2$ with a winding number $\ql = -1$. The difference between particles and antiparticles is associated with orientation on circles in phase space and orientation on the time axis. 

To describe quantum harmonic oscillators, it is first necessary to define a principal bundle $P(\R^2, \sU(1)_{\sf v})$ with structure group $\sU(1)_{\sf v}$ and connection $\Av$. To do this, consider the space
\begin{equation}\label{3.1}
P(\R^2, \sU(1)_{\sf v})=\R^2\times S^1\with S^1\cong\sU(1)_{\sf v}
\end{equation}
and introduce on it a one-form $\Av=\CA_x\dd x + \CA_p\dd p$ along $\R^2$ whose components are vector fields along $S^1$,
\begin{equation}\label{3.2}
\CA_x = A_x\dpar_\theta \ ,\quad \CA_p = A_p\dpar_\theta \ ,\quad A=A_x\dd x + A_p\dd p= \theta_{\R^2}=\sfrac12\,p\dd x - \sfrac12\,x\dd p\ .
\end{equation}
Here $0\le\theta <2\pi$ is a coordinate on $S^1$. Then we introduce on the space \eqref{3.1} vector fields
\begin{equation}\label{3.3}
\nabla_x = \dpar_x + A_x\dpar_\theta\ ,\quad \nabla_p = \dpar_p+ A_p\dpar_\theta\und\nabla_\theta =\dpar_\theta
\end{equation}
and dual one-forms
\begin{equation}\label{3.4}
\Xi^x=\dd x\ ,\quad \Xi^p=\dd p\und \Xi^\theta = \dd\theta - A_x\dd x - A_p\dd p\ .
\end{equation}
Vector fields \eqref{3.3} and one-forms \eqref{3.4} form a frame and coframe on $P(\R^2, \sU(1)_{\sf v})$ as a manifold.

Calculating commutator of vector fields \eqref{3.3}, we obtain the curvature
\begin{equation}\label{3.5}
\CF_{xp}=[\nabla_x, \nabla_p]=F_{xp}\dpar_\theta = \omega_{xp}\dpar_\theta =- \dpar_\theta
\end{equation}
of the connection $\Av$. Note that fields interacting with $\Av$ depend on $\theta$ as $\exp (\im\qv\theta)$, $\qv\in \Z$, and for the cases of quantum charges $\qv=\pm 1$ we are considering here, we obtain
\begin{equation}\label{3.6}
\nabla_x = \dpar_x \pm \im A_x\ ,\quad \nabla_p = \dpar_p\pm \im\,A_p\und\CF_{xp} =\pm\im\omega_{xp}=\mp\im\ .
\end{equation}
We call the number $\qv$ the quantum charge; it distinguishes between particles ($\qv\ge 0$) and antiparticles ($\qv\le 0$), with $\qv =0$ corresponding to neutral particles. Quantum charge $\qv$ is related to the winding number \eqref{1.1}. In \eqref{3.6} we have $\Av =\im\theta^{}_{\R^2}$ for $\qv =1$ and $\Av =(\im\theta^{}_{\R^2})^*= - \im\theta^{}_{\R^2}$ for $\qv =-1$. 

\medskip

\noindent  {\bf Complex line bundle $L_\C^\pm$}. The fibres of the principal bundle \eqref{3.1} over points $x\in\R^2$ are groups $\sU(1)_{\sf v}$. Let us consider complex one-dimensional vector spaces $V^\pm$ on which the group $\sU(1)_{\sf v}$ acts according to the rule $V^\pm\ni\psi_\pm\mapsto\exp (\pm\im\theta )\psi_\pm\in V^\pm$. We associate with $P(\R^2, \sU(1)_{\sf v})$ the complex line bundles
\begin{equation}\label{3.7}
L_\C^\pm = P\times_{\sU(1)_{\sf v}}V^\pm =\left\{ P\times V^\pm\ni (p, \psi_\pm )\sim (pg_\pm^{-1}, g^{}_\pm\psi_\pm )\in P\times V^\pm\right\}=\R^2\times V^\pm\ ,
\end{equation}
where $g^{}_\pm =\exp (\pm\im\theta )$.  The sign ``$\sim$" means equivalence under the action of the group $\sU(1)_{\sf v}$ on the direct product $P\times V^{\pm}$ of the space $P$ and $V^\pm$.

Spaces $V^\pm$ are introduced as follows. Let us consider two-dimensional columns
\begin{equation}\label{3.8}
\Psi = \begin{pmatrix}\psi^1\\\psi^2\end{pmatrix}\in\C^2
\end{equation}
with complex components $\psi^1, \psi^2$. These columns are acted upon by matrix
\begin{equation}\label{3.9}
J=\begin{pmatrix}0&-1\\1&0\end{pmatrix}\ ,
\end{equation}
which is the generator of group $\sU(1)_{\sf v}$. In the space of $\C^2$-vectors \eqref{3.8} we introduce a basis of eigenvectors of the matrix $J$:
\begin{equation}\label{3.10}
Jv^{}_\pm =\pm\im v^{}_\pm \ \Rightarrow\ v^{}_\pm  =\frac{1}{\sqrt 2} \begin{pmatrix}1\\ \mp\im\end{pmatrix},\quad
v_-=v_+^* \ ,\quad v_\pm^\+v^{}_\pm =0\ ,\quad     v_\pm^\+v^{}_\mp =0\ ,
\end{equation}
where ``$*$" means complex conjugation. These vectors $v^{}_\pm$ are basis vectors in the spaces $V^\pm$, i.e. $\C^2 = V^+\oplus V^- =\C\oplus\bar\C$. Now vector \eqref{3.8} can be expanded in $V^\pm$-parts,
\begin{equation}\label{3.11}
\Psi = \begin{pmatrix}\psi^1\\ \psi^2\end{pmatrix}=\Psi_+ +\Psi_- = \psi_+ v_+ + \psi_- v_-\in V^+\oplus V^-\with \psi_\pm = \frac{1}{\sqrt 2}\, (\psi^1\pm\im\psi^2)\ .
\end{equation}
These $\psi_\pm^{}$ are complex coordinates on fibres $V^\pm$ of the bundles $L_\C^\pm$ in \eqref{3.7}. Note that $\psi^1, \psi^2$ are complex, therefore in the general case $\psi_-$ is not complex conjugate to $\psi_+$ despite the fact that $v_-=\bar v_+\equiv v_+^*$.

We introduce a Hermitian structure on the bundles $L_\C^\pm$ by equipping fibres $V^\pm$ with the Hermitian inner product
\begin{equation}\label{3.12}
\langle\psi_\pm^{} , \psi_\pm^{}\rangle =\Psi_\pm^\+\Psi_\pm^{}=\psi_\pm^* \psi_\pm
\end{equation}
It is obvious that the metric \eqref{3.12} is invariant under the action $\psi_\pm^{}\mapsto g_\pm^{}\psi_\pm^{}$ of the group $\sU(1)_{\sf v}$ with
$g_\pm^{}=\exp(\pm\im\theta )\in \sU(1)_{\sf v}$.

\medskip

\noindent  {\bf Complex vector bundle $L_{\C^2}$}. Group $\sU(1)_{\sf v}$ acts on the space of $\C^2$-vectors \eqref{3.8} by multiplying on the left by the matrix
\begin{equation}\label{3.13}
e^{\theta J}=\cos\theta + J\sin\theta =  \begin{pmatrix}\cos\theta&-\sin\theta\\ \sin\theta&\cos\theta\end{pmatrix}\in \sSO(2)_{\sf v}\ .
\end{equation}
For subspaces $V^\pm$ in $\C^2=V^+\oplus V^-$ we obtain 
\begin{equation}\label{3.14}
\Psi (\theta)=e^{\theta J}\Psi =\Psi_+(\theta) + \Psi_-(\theta )=e^{\im\theta}\psi_+v_+ + e^{-\im\theta}\psi_-v_-\ ,
\end{equation}
which coincides with definition \eqref{3.7} of spaces $V^\pm$ in $L_\C^\pm$. Note that the action of the generator $\dpar_\theta$ of group  $\sU(1)_{\sf v}$ on $\Psi (\theta)$ has the form
\begin{equation}\label{3.15}
\dpar_\theta\Psi (\theta)=J\Psi (\theta)=\im\Psi_+ (\theta)-\im\Psi_- (\theta)\ ,
\end{equation}
i.e. it is equivalent to the action of the generator $J$ from  \eqref{3.9}.

The column vectors \eqref{3.11} are sections of the complex vector bundle
\begin{equation}\label{3.16}
L_{\C^2}:=L_{\C}^+\oplus L_{\C}^-
\end{equation}
with the structure group given in \eqref{3.13}-\eqref{3.15}. The $\C^2$-bundle \eqref{3.16} inherits its connection $\Av$ and curvature $\Fv$ from connections and curvature \eqref{3.6} on $L_\C^\pm$,
\begin{equation}\label{3.17}
\Av = AJ=\theta_{\R^2}J\und\Fv = \dd\Av =J\dd\theta_{\R^2}=\omega_{\R^2}J\ .
\end{equation}
The components of this connection are given in \eqref{3.2}:
\begin{equation}\label{3.18}
A_x =\sfrac12 p\und A_p =-\sfrac12 x\ .
\end{equation}
Accordingly, the covariant derivatives on $L_{\C^2}$ have the form
\begin{equation}\label{3.19}
\nabla_x = \dpar_x  + A_xJ = \dpar_x + \sfrac12 pJ \und \nabla_p = \dpar_p  + A_pJ = \dpar_p - \sfrac12 xJ \ .
\end{equation}
Note that $\Av\in \rso(2)_{\sf v}$ and therefore the connection $\Av$ is compatible with the Hermitian metric
\begin{equation}\label{3.20}
\langle\Psi , \Psi\rangle =\Psi^\+\Psi
\end{equation}
on $L_{\C^2}$.

\medskip

\noindent  {\bf Operators $\ph$ and $\xh$.} We consider quantum mechanics as a gauge theory of fields $\psi_\pm\in\Gamma (L_\C^\pm)$ with $\qv=\pm 1$ interacting with gauge fields $\Av =\im\qv\theta_{\R^2}$ defined on these bundles. We also use the bundle \eqref{3.16} to describe $\psi_\pm$ simultaneously as sections \eqref{3.11}, \eqref{3.14} of the bundle $L_{\C^2}$. 

As has been noted more than once, the spaces $\Gamma(L_\C^\pm)$ of sections of bundles $L_\C^\pm$ are too large and they need to be narrowed down to spaces of irreducible representations of CCR by imposing conditions
\begin{equation}\label{3.21}
X_\pm\psi_\pm = 0\ ,\quad \psi_\pm\in\Gamma(L_\C^\pm)\ ,\quad X_\pm\in\Gamma (\CT_\pm )\ ,\quad \CT_\pm\subset T^\C\R^2\ ,
\end{equation} 
where $X_\pm$ are vector fields from the subbundles $\CT_\pm$ of the complexified tangent bundle of the phase space $T^*\R$ of oscillators. For real polarizations $\CT_\pm$, they are real subbundles of the tangent bundle $T\R^2$ and we can consider polarization for sections of the bundle $L_{\C^2}=L_\C^+\oplus L_\C^-$ since $\CT_+=\CT_-$.

In the two-dimensional case we are considering, the real polarization is either the independence of sections $\Psi \in \Gamma (L_{\C^2}$) from the momenta,
\begin{equation}\label{3.22}
\dpar_p\Psi =0\quad\Rightarrow\quad\Psi =\Psi (x,t)\ ,
\end{equation}
or their independence from the coordinates, 
\begin{equation}\label{3.23}
\dpar_x\Psi =0\quad\Rightarrow\quad\Psi =\Psi (p,t)\ .
\end{equation}
After imposing one of these conditions, we arrive at quantum mechanics in coordinate or momentum representation.

Let us see how this works for polarization \eqref{3.22}. Note that vector field $\dpar_p$ does not commute with covariant derivatives \eqref{3.19}, which is unacceptable. However, connection \eqref{3.18} can be transformed using the action of the group $\CG$ of unitary automorphisms of the bundle $L_{\C^2}$,
\begin{equation}\label{3.24}
\CG = C^\infty (\R^2, \sU(1)_{\sf{v}})\ ,
\end{equation}
with elements $g=\exp(\alpha(x,p)J)$. Here $\alpha(x,p)$ is a real function on $\R^2$. If we choose $\alpha =-\sfrac12 px$, we get
\begin{equation}\label{3.25}
A_x^\alpha =A_x + g^{-1}\dpar_x g=0\ ,\quad A_p^\alpha =A_p + g^{-1}\dpar_p g=-x
\end{equation}
\begin{equation}\label{3.26}
\Rightarrow\ \nabla_x^\alpha =\dpar_x\und \nabla_p^\alpha =\dpar_p - xJ\ .
\end{equation}
Now the covariant derivatives commute with the derivative $\dpar_p$ in  \eqref{3.22} and we can introduce the operators
\begin{equation}\label{3.27}
\xh := J\nabla_p^\alpha = x + J\dpar_p\und \ph := -\im\nabla_x^\alpha = -\im\dpar_x\ ,
\end{equation}
which are the standard operators of coordinate and momentum when acting on polarized sections  \eqref{3.22}. Thus, operators $\ph$ and $\xh$ are the covariant derivatives  \eqref{3.27} in the bundles $L_\C^\pm$ and the canonical commutation relation (CCR) is
\begin{equation}\label{3.28}
[\ph , \xh]=-\im J[\nabla_x^\alpha , \nabla_p^\alpha ] =-\im J\CF_{xp}=-\im\ .
\end{equation}
This is nothing more then the curvature $\Fv$  multiplied by $-\im J=\pm 1$ on $L_\C^\pm$.

Note that the CCR \eqref{3.28} does not depend  on the choice of function $\alpha(x,p)$. The curvature $\Fv$ of the connection $\Av$ on the bundle 
$L_{\C^2}$ defines both the CCR \eqref{3.28} and the uncertainty relation. Recall that $\Av$ and $\Fv$ are vacuum gauge fields and the field $\Av$ defines the potential energy of vacuum through the covariant Laplacian $g^{pp}\nabla_p^\alpha\nabla_p^\alpha$ along the momentum space. A more general potential energy $V(x)$ can be introduced either as a function of the covariant derivative $\nabla_p^\alpha$ or through the component $g^{pp}$ of the metric on phase space.

To use polarization \eqref{3.23}, one should apply the automorphism $g^{-1}=\exp(-\alpha J)=\exp(\sfrac12 pxJ)$, obtaining
\begin{equation}\label{3.29}
\nabla_x^{-\alpha}=\dpar_x + pJ\ ,\quad \nabla_p^{-\alpha}=\dpar_p\quad\Rightarrow\quad\ph=-\im p J -\im\dpar_x\und\xh =J\dpar_p\ .
\end{equation}
Then in the momentum representation we obtain
\begin{equation}\label{3.30}
\ph =p\und\xh =\im\dpar_p\ \mbox{on}\ L_\C^+\quad(\mbox{particles})
\end{equation}
\begin{equation}\label{3.31}
\ph =-p\und\xh =-\im\dpar_p\ \mbox{on}\ L_\C^-\quad(\mbox{antiparticles})
\end{equation}
Note that \eqref{3.30} corresponds to the standard definition, and \eqref{3.31} reflects the fact that for antiparticles we have $p\mapsto -p$, as discussed earlier. The CCR \eqref{3.28} does not change.

In conclusion of this section, we note that for the automorphism generated by the element $h=\exp(px_0J)$ applied to \eqref{3.27}, we obtain translations
\begin{equation}\label{3.32}
\xh\ \mapsto\ \xh = x-x_0+J\dpar_p\und \ph\mapsto\ph =-\im\dpar_x\ ,
\end{equation}
and, therefore, coherent states can also be easily described within the framework of the approach under consideration.

\section{Complex polarizations}


\noindent  {\bf Dolbeault operators.} We considered bundles $L_\C^\pm$ with anti-Hermitian connections $\Av$ and automorphisms \eqref{3.24}-\eqref{3.26}, \eqref{3.29} that transform covariant derivatives with this connections into operators $\xh$ and $\ph$ in irreducible coordinate representation \eqref{3.27} and momentum representation \eqref{3.29}-\eqref{3.31}. Now we will describe how complex polarizations $\CT_\pm\subset T^\C\R^2$ are introduced. For them, particles $\Psi_+\in L_\C^+$ are holomorphic functions of the complex coordinate $z_+$ from \eqref{2.7} (Segal-Bargmann representation \cite{Segal, Bar, Hall}) and antiparticles $\Psi_-\in L_\C^-$ are holomorphic functions of the complex coordinate $z_-$ on $\R^2$ given in \eqref{2.15} (antiholomorphic in $z_+$ since $z_-=\zb_+$). 

To define holomorphic structures in the bundles $L_\C^\pm$, we introduce the Dolbeault operators
\begin{equation}\label{4.1}
\bar\dpar_{L_\C^\pm}^{}=\dd\zb_\pm\left (\frac{\dpar}{\dpar\zb_\pm }+\frac{z_\pm}{2w^2}\right )\ ,
\end{equation}
and impose the conditions
\begin{equation}\label{4.2}
\bar\dpar_{L_\C^\pm}^{}\Psi^{}_\pm =0
\end{equation}
on sections $\Psi^{}_\pm$ of the bundles $L_\C^\pm$. These are conditions for holomorphic polarization and their solutions are functions
\begin{equation}\label{4.3}
\Psi^{}_\pm =\psi_\pm (z_\pm , t)v_\pm^c\with v_\pm^c=e^{-z\zb/2w^2} v_\pm\ ,
\end{equation}
where $z\zb = z_+\zb_+=z_-\zb_-=\sfrac12\,(x^2 + w^4p^2)$. Note that $z_-=\zb_+$ and the operator $\bar\dpar_{L_\C^-}^{}$ is complex conjugate to the operator  $\bar\dpar_{L_\C^+}^{}$, but the functions $\psi_+(z_+, t)$ and $\psi_-(z_-, t)$ in the general case are not related by complex conjugation.

\medskip

\noindent  {\bf Covariant derivatives.} Note that the basis vectors $v_\pm$ of the complex line bundles $L_\C^\pm$ define the Hermitian metrics \eqref{3.12} and \eqref{3.20} in fibres. These are Hermitian bases of Hermitian bundles. At the same time, the basis vectors $v_\pm^c$ in \eqref{4.3} define in $L_\C^\pm$ complex bases associated with the principal bundle
\begin{equation}\label{4.4}
P(\R^2, \sGL(1, \C)_{\sf v})\to \R^2
\end{equation}
having the structure group GL$(1, \C)_{\sf v} =\C^*\supset\sU(1)_{\sf v}$, and the previously considered bundle $P(\R^2, \sU(1)_{\sf v})$ is a Hermitian subbundle \cite{KobNom} in the bundle \eqref{4.4}.  The function
\begin{equation}\label{4.5}
\phi_0=\exp\Bigl(-\frac{z\zb}{2w^2}\Bigr)
\end{equation}
in \eqref{4.3} is an element of the group GL$(1,\C)_{\sf v}$ that defines a mapping of Hermitian bases $v_\pm$ into holomorphic bases $v_\pm^c$ along which the holomorphic sections \eqref{4.3} of the bundles $L_\C^\pm$ are decomposed. 

In covariant derivatives \eqref{3.6} we initially use anti-Hermitian connections $\Av=\pm\im\theta^{}_{\R^2}$. In complex coordinates $z_\pm$ we have
\begin{equation}\label{4.6}
\nabla^{}_{z_\pm}=\frac{1}{\sqrt 2}\left (\nabla_x\pm \frac{\im}{w^2}\,\nabla_p\right )=\dpar^{}_{z_\pm}+\frac{\zb_\pm}{2w^2}\ ,\quad
\nabla^{}_{\zb_\pm}=\frac{1}{\sqrt 2}\left (\nabla_x\mp \frac{\im}{w^2}\,\nabla_p\right )=\dpar^{}_{\zb_\pm}-\frac{z_\pm}{2w^2}
\end{equation}
which implies that
\begin{equation}\label{4.7}
A^{}_{z_\pm}=\frac{\zb_\pm}{2w^2}\und A^{}_{\zb_\pm}=-\frac{z_\pm}{2w^2}\ .
\end{equation}
Using function \eqref{4.5} as an automorphism of the bundle $L_\C^\pm$, we obtain the following components of the connection in the holomorphic bases $v^c_\pm$:
\begin{equation}\label{4.8}
A^{\phi_0}_{z_\pm}=A^{}_{z_\pm}+\phi_0^{-1}\dpar^{}_{z_\pm}\phi_0=0\ ,\quad  A^{\phi_0}_{\zb_\pm}=A^{}_{\zb_\pm}+\phi_0^{-1}\dpar^{}_{\zb_\pm}\phi_0=-\frac{z_\pm}{w^2}
\end{equation}
\begin{equation}\label{4.9}
\Rightarrow\ \nabla^{\phi_0}_{z_\pm}=\frac{\dpar}{\dpar z_\pm}\und \nabla^{\phi_0}_{\zb_\pm}=\frac{\dpar}{\dpar \zb_\pm}-\frac{z_\pm}{w^2}\ .
\end{equation}
Under this automorphism, the Dolbeault operators \eqref{4.1} transform into ordinary $\bar\dpar$-operators
\begin{equation}\label{4.10}
\bar\dpar_{L_\C^\pm}^{\phi_0}=\dd\zb_\pm\dpar^{}_{\zb_\pm}
\end{equation}
with partial derivatives $\dpar^{}_{\zb_\pm}$. They commute with covariant derivatives \eqref{4.9}. Therefore, connections \eqref{4.8} are holomorphic but not Hermitian.

\medskip

\noindent  {\bf Ladder operators.} From \eqref{4.6} we see that the annihilation and creation operators for bundles $L_\C^\pm$ have the form
\begin{equation}\label{4.11}
a^{}_\pm = w\nabla_{z_\pm}\und a_\pm^\+ = -w\nabla_{\zb_\pm}\with [a^{}_\pm ,  a_\pm^\+]=1
\end{equation}
when they act on sections $\Psi_\pm$ from \eqref{4.3}. After transformations \eqref{4.8}-\eqref{4.10} they take the usual form
\begin{equation}\label{4.12}
a^{}_\pm =\frac{\dpar}{\dpar z_\pm^\prime}\und a_\pm^\+ =  z_\pm^\prime\quad\mbox{for}\quad  z_\pm^\prime :=\frac{z_\pm}{w}
\end{equation}
when acting on holomorphic functions $\psi_\pm (z_\pm , t)$ of $z_\pm$. The scalar product of such functions has the form
\begin{equation}\label{4.13}
\Psi^{\+}_\pm\Psi^{}_\pm = = \psi_\pm^*(z_\pm , t)\psi_\pm(z_\pm , t)\exp\left(-\frac{z_\pm\zb_\pm}{2w^2}\right)\ ,
\end{equation}
as it should be in the Segal-Bargmann representation \cite{Segal, Bar, Hall}.

\medskip

\noindent  {\bf Complex $\ \Rightarrow\ $ real.} Note that real polarization \eqref{3.22} and transformations \eqref{3.24}-\eqref{3.26} to the coordinate representation can be obtained as the limiting case of complex polarization \eqref{4.2}:
\begin{equation}\label{4.14}
\mathop{\lim}_{w\to 0}\left(w^2\bar\dpar_{L_\C^\pm}^{}\Psi_\pm\right)=0\ \Rightarrow\ \bigl(\dpar_p\pm\sfrac{\im x}{2}\bigr)\Psi_\pm (x,p,t)=0\ \Rightarrow\ \Psi_\pm =e^{\mp\im px/2}\psi_\pm(x,t)v_\pm\ .
\end{equation}
It is not difficult to verify that operators \eqref{4.11} on such functions reduce to operators
\begin{equation}\label{4.15}
a^{}_\pm = w\nabla_{z_\pm}^\alpha = w e^{\pm\im px/2}\circ \nabla_{z_\pm}\circ e^{\mp\im px/2}=\frac{w}{\sqrt 2}\Bigl(\dpar_x + \frac{ x}{w^2}\Bigr)\ ,
\end{equation}
\begin{equation}\label{4.16}
a^{\+}_\pm = -w\nabla_{\zb_\pm}^\alpha = -w e^{\pm\im px/2}\circ \nabla_{\zb_\pm}\circ e^{\mp\im px/2}=\frac{w}{\sqrt 2}\Bigl( \frac{ x}{w^2}-\dpar_x \Bigr)\ ,
\end{equation}
i.e. to standard ladder operators in coordinate representation. Similarly, the real polarization \eqref{3.23} and the transformation \eqref{3.29}-\eqref{3.31} to the momentum representation can be obtained as another limit of the complex polarization \eqref{4.2}:
\begin{equation}\label{4.17}
\mathop{\lim}_{w\to \infty}\bigl(w^{-2}\bar\dpar_{L_\C^\pm}^{}\Psi_\pm\bigr)=0\ \Rightarrow\ \bigl(\dpar_x\mp\sfrac{\im p}{2}\bigr)\Psi_\pm (x,p,t)=0\ \Rightarrow\ \Psi_\pm =e^{\pm\im px/2}\psi_\pm(p,t)v_\pm\ .
\end{equation}
Operators \eqref{4.11} on  functions \eqref{4.17} are reduced to ladder operators $a_\pm$, $a_\pm^\+$ in momentum representation. We will not write out their explicit form.

\medskip

\noindent  {\bf Covariant Laplacians.} Having considered real and complex polarizations, we will move on to defining Hamiltonians for harmonic oscillators with $\qv =\pm 1$. To do this, we introduce covariant Laplacians,
\begin{equation}\label{4.18}
\Delta_2^\pm = \nabla_{z_\pm}\nabla_{\zb_\pm}+\nabla_{\zb_\pm}\nabla_{z_\pm}\ ,
\end{equation}
acting on polarized sections \eqref{4.3} of the bundles $L_\C^\pm$ . Substituting the explicit form \eqref{4.6} of covariant derivatives into \eqref{4.18}, we obtain
 \begin{equation}\label{4.19}
\Delta_2^\pm \Psi_\pm^{}= -\frac{2}{w^2}\biggl[\Bigl(z_\pm\frac{\dpar}{\dpar z_\pm}+ \frac12\Bigr)\psi_\pm (z_\pm , t)\biggr]v_\pm^c\ .
\end{equation}
We can now introduce natural geometric Hamiltonians
 \begin{equation}\label{4.20}
\hat H_\pm =-\frac{1}{2m}\Delta_2^\pm =\omega \Bigl(z_\pm\frac{\dpar}{\dpar z_\pm}+ \frac12\Bigr)
\end{equation}
defined on functions $\psi_\pm (z_\pm , t)$. Here we used the connection \eqref{2.9} between $w^2$ and $m\omega$.

\medskip

\noindent  {\bf Schr\"odinger equations.} Recall that $z_-=\zb_+$ and $\Psi_\pm$ are sections of complex conjugate bundles $L_\C^\pm$. Therefore, the Schr\"odinger equations for them have the form
 \begin{equation}\label{4.21}
-\im\,{\dpar_t\psi_+}(z,t)=\omega (z\dpar_z +\sfrac12)\psi_+(z,t)\ ,
\end{equation}
 \begin{equation}\label{4.22}
\im\,\dpar_t\psi_-(\zb,t)=\omega (\zb\dpar_\zb +\sfrac12)\psi_-(\zb,t)\ ,
\end{equation}
where $\dpar_t:=\dpar/\dpar t$.
Recall that in Section 2 we choose $e^{\im\omega t}$ and $e^{-\im\omega t}$ as positive and negative frequencies to match the signs of the winding numbers. This is why, in \eqref{4.21} the operator $-\im\dpar_t$ (and not $\im\dpar_t$) is used to decompose the space $\CH^{(1)}=L^2(S^1, \C)$ into the direct sum $\CH^{(1)}_+\oplus\CH^{(1)}_-$ of positive and negative subspaces (and similarly for $L^2(\R, \C)$) \cite{PS}.
This choice, like the distinction between particles and antiparticles, is tightly related to the choice of {\it orientation} in spaces $S^1$ and $\R$.

Equations \eqref{4.21} and \eqref{4.22} can be combined into one equation for sections
 \begin{equation}\label{4.23}
\Psi=\psi_+(z,t) v_+ + \psi_- (\zb , t)v_-
\end{equation}
of the bundle $L_{\C^2}^{}=L_{\C}^{+}\oplus L_{\C}^{-}$, obtaining 
 \begin{equation}\label{4.24}
\dpar_t\Psi =\omega (\CJ +\sfrac12 J)\Psi\ .
\end{equation}
Here $\CJ$ is the vector field  \eqref{2.19} on the base $\R^2$ of the bundle $L_{\C^2}\to\R^2$ and it is the generator of the group U(1)$_l$ of rotations on $\R^2$, and $J$ is the generator of the group U(1)$_{\sf v}$ of rotations $\psi_\pm\mapsto e^{\pm\im\theta}\psi_\pm$ of the coordinates $\psi_\pm$ on fibres of bundles $L_\C^\pm\to\R^2$. This generator can be represented by a vector field or matrix (see \eqref{3.15}):
 \begin{equation}\label{4.25}
J=\im\psi_+\frac{\dpar}{\dpar\psi_+}-\im\psi_-\frac{\dpar}{\dpar\psi_-}=\dpar_\theta\quad\Leftrightarrow\quad 
J=\begin{pmatrix}0&-1\\1&0\end{pmatrix}\ \mbox{on}\ \Psi(\theta )\ .
\end{equation}
Writing in the form  \eqref{4.24} emphasizes and clarifies the U(1)-nature of the Schr\"odinger equation. Equation  \eqref{4.24} can also be rewritten as
 \begin{equation}\label{4.26}
-\im\dpar_t\Psi =\omega (Q_l +\sfrac12\,Q_{\sf v})\Psi\ ,
\end{equation}
where
 \begin{equation}\label{4.27}
Q_l:=-\im\CJ = z\dpar_z - \zb\dpar_\zb\und Q_{\sf v}=-\im J
\end{equation}
are the winding number operators in the base $\R^2$ and fibres $\C^2=\C\oplus\bar\C$ of the bundle $L_{\C^2}$. In the case we are considering, the eigenvalues of the operator $Q_{\sf v}$ are fixed at $\qv=\pm 1$, and the eigenvalues of the operator $Q_l$ can be any integers.

\medskip

\noindent  {\bf Classical and quantum: comparison}. To compare the Schr\"odinger equation \eqref{4.24} for quantum oscillator with equations \eqref{2.21} for classical oscillators, we introduce the vector
 \begin{equation}\label{4.28}
Z=z_+v_+ + z_-v_-\ .
\end{equation}
When using $Z$, equations \eqref{2.21} are combined into one equation
 \begin{equation}\label{4.29}
\dpar_tZ= \omega\CJ Z\ ,
\end{equation}
which can be compared with equation \eqref{4.24}. Comparing the solutions of these equations, we have
\begin{equation}\label{4.30}
\begin{array}{ll}
z_0, \ql =1\quad \to\quad (z, \psi^+_n(z)),\ \ql =n\ge 0, \  \qv=1\\[2pt]
\zb^\prime_0, \ql =-1\quad \to\quad (\zb, \psi^-_n(\zb)),\ \ql =-n\le 0, \  \qv=-1\ ,
\end{array}
\end{equation}
where $\ql$ and $\qv$ are winding numbers. In addition, classical oscillator is a point moving in a circle on the plane $\R^2$, and quantum oscillator is a Riemann surface $\C/\Z_n$ in $L_\C^+$ (or $\bar\C/\Z_n$ in $L_\C^-$), each point of which moves in a circle in $L_\C^+$ (or $L_\C^-$), $n=0,1,...$. The Riemann surface  $\C/\Z_n$ will be described in the next section.

\section{Harmonic oscillators and orbifolds}

\noindent  {\bf Solutions}. We will discuss only solutions $\psi_+$ of equations \eqref{4.21} in the Segal-Bargmann representation, since for $\psi_-$ everything is similar. We will also move on to dimensionless coordinate $z^\prime =z/w$ and omit the prime in $z^\prime$ in the formulae below.

We have a discrete set of solutions
\begin{equation}\label{5.1}
\Psi_+(n) =(\pi n!)^{-\frac12} \bigl(e^{\im\omega t}z\bigr)^nv_+^c(t)\ , \quad v_+^c(t)=e^{\im\omega t/2}v_+^c=e^{-z\zb/2}e^{\im\omega t/2}v_+
\end{equation}
with the energy\footnote{We return Planck's constant to the expressions for the energy levels.}
\begin{equation}\label{5.2}
E_n=\hbar\omega (\ql + \sfrac12\qv) = \hbar\omega (n+\sfrac12)=\hbar\omega n + \sfrac12\,\hbar\omega\ ,\ n=0,1,...,
\end{equation}
where the term $\hbar\omega n$ is the energy of rotating surface
\begin{equation}\label{5.3}
\C/\Z_n=\Bigl\{(z, \psi_n (z) =z^n)\Bigr\}\subset L_\C^+\ ,
\end{equation}
and the term $ \sfrac12\,\hbar\omega$ is the energy of the rotating basis vector $v_+^c$ of fibres $\C$ of the bundle $L_\C^+$. 

Note that the squared modulus of the function \eqref{5.1} is the Husimi $Q$-function,
\begin{equation}\label{5.4}
Q_n(x, p) = \Psi_+^\+(n)\Psi_+^{}(n)=\frac{1}{\pi n!}\,(z\zb)^n e^{-z\zb}\ ,
\end{equation}
which is a quasiprobability distribution in phase space. From the point of view of gauge theory $(L_\C^+, \Av , \Psi_+^{})$, this function is the quantum charge density of  section $\Psi_+^{}(n)$ of the bundle $L_\C^+\to \R^2$. 

\medskip

\noindent  {\bf Cross sections of $L_\C^+$}. So, we consider the bundle $L_\C^+$ with projection $\pi$,
\begin{equation}\label{5.5}
\pi: \ L_\C^+=\R^2\times\C\cong\C^2\ \stackrel{\C}{\to}\ \R^2\cong\C\ ,
\end{equation}
and solutions \eqref{5.1} to equations \eqref{4.21}. Recall that the graph of a function $f: X\to Y$ can be identified with a function $\sigma$ taking value in the Cartesian product,
\begin{equation}\label{5.6}
\sigma :\ X\to X\times Y\ ,\quad \sigma (x)=(x, f(x))\in X\times Y\ .
\end{equation}
In our case, we consider a section $\sigma_n$ of the bundle \eqref{5.5},
\begin{equation}\label{5.7}
\sigma_n :\ \C\to \C^2=  L_\C^+\ ,\quad   \sigma_n (z)=(z, \psi_n(z))=(z, z^n)\ ,
\end{equation}
included in the solution \eqref{5.1}, where the function
\begin{equation}\label{5.8}
\psi_n: z\to z^n
\end{equation}
defines the graph in $L_\C^+$. This graph is a one-dimensional complex surface $\C/\Z_n$ in $L_\C^+$ and solution \eqref{5.1} describes a standing wave on this surface.

\medskip

\noindent  {\bf Orbifold $\C/\Z_n$}. Recall that the map \eqref{5.8} with $n\ge 2$ is the branched covering of degree $n$, where $z=0$ is the branch point. For $n=1$ the map $\psi_1$ is the identity. The surface \eqref{5.7} in $L_\C^+$ is the orbifold $\C/\Z_n$, where $\Z_n=\Z/n\Z$ is the cyclic group of order $n$, generated by an element $\zeta$ with $\zeta^n=1$, i.e. $\zeta$ is $n$-th root of unity. Thus, we have the projection
\begin{equation}\label{5.9}
\psi_n: \ \C\to \C/\Z_n\ ,
\end{equation}
and $\C$ is the total space of this bundle. The function \eqref{5.8} is an ordinary function and its inverse
\begin{equation}\label{5.10}
\psi_n^{-1}:\ \C/\Z_n\to \C
\end{equation}
is a multivalued function
\begin{equation}\label{5.11}
z=\psi_n^{1/n}\ ,
\end{equation}
where, abusing the notation, we denoted the complex coordinate on $\C/\Z_n$ by $\psi_n$. 

\medskip

\noindent  {\bf $\C/\Z_n$ as a cone.} The orbifold $\C/\Z_n$ is a metric cone of $S^1/\Z_n$. Recall that the metric cone over the circle $S^1$ is Euclidean space $C(S^1)=\C{\setminus}\{0\}$ and hence may in fact be continued non-singularly at the cone tip $z=0$. For $n\ge 2$ ($\Z_1=\Id$) the metric cone $C(S^1/\Z_n)$ is singular at the origin $z=0$, since the action of $\sU(1)\supset\Z_n$ is free only except at the origin. The group $\Z_n$ acts on $\R^2\cong\C$ by a counterclockwise rotations through the angle $2\pi/n$ about the origin and the quotient is a cone with the cone angle $2\pi/n$:
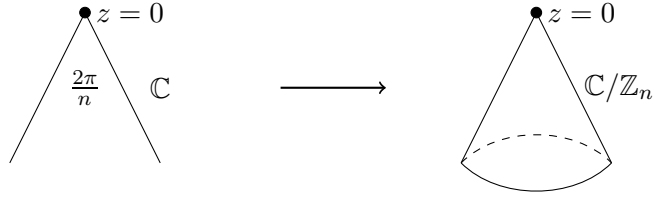
\begin{center}
\begin{tikzpicture}
\draw (0,0) -- (-1,-2);
\draw (0,0) -- (1,-2);
\filldraw[black] (0,0) circle (2pt);
\node at  (0.6,0) {$z=0$};
\node at  (0,-1) {$\frac{2\pi}{n}$};
\node at  (1,-1) {$\mathbb{C}$};
\draw[->, thick] (2.6,-1) -- (4,-1); 
\draw (6,0) -- (5,-2);
\draw (6,0) -- (7,-2);
\filldraw[black] (6,0) circle (2pt);
\node at  (6.6,0) {$z=0$};
\node at  (7.1,-1) {$\mathbb{C}/\mathbb{Z}_n$};
\draw (5,-2) .. controls (5.5,-2.5) and (6.5,-2.5) .. (7,-2);
\draw[dashed](5,-2) .. controls (5.5,-1.5) and (6.5,-1.5) .. (7,-2);
\end{tikzpicture}
\end{center}
Note that on $\C{\setminus}\{0\}$ in mapping \eqref{5.9} there are $n$ different points $z_i=\zeta^iz\in\C$, $i=0,...,n-1$, mapped to the same point $\psi_n=z^n$ on $\C/\Z_n$, where  $\zeta =\exp (2\pi\im/n)$. The action of group $\Z_n$ on $\C$ defines an equivalence relation and the part of the plane $\C$, which is cut out by rays with angle $\frac{2\pi}{n}$, is a representative of the cone $\C/\Z_n$.

\medskip

\noindent  {\bf Metric and curvature of $\C/\Z_n$.} Metric on $\C/\Z_n$ is induced from the metric on $\C$. Using the connection \eqref{5.11} between the coordinates on $\C$ and $\C/\Z_n$, we obtain
\begin{equation}\label{5.12}
\dd s^2_{\C/\Z_n} = 2\dd z\dd\zb|_{\C/\Z_n}^{} = 2(\bar\psi_n\psi_n)^{\frac{1-n}n}\dd\psi_n\dd\bar\psi_n =\dd\rho^2 + \frac{\rho^2}{n^2}\dd\vph_n^2\ ,
\end{equation}
where
\begin{equation}\label{5.13}
\sqrt 2\, z=\rho\exp(\im\vph )\ ,\quad \sqrt 2\, \psi_n=\rho_n\exp(\im\vph_n )\ ,\quad \rho_n=\rho^n\und 0\le\vph_n<2\pi\ .
\end{equation}
The Levi-Civita connection $\Gamma_n$ of the metric \eqref{5.12} on the cone $\C/\Z_n$ can be written in the form
\begin{equation}\label{5.14}
\nabla^{}_{\Gamma_n}=\dd\psi_n\frac{\dpar}{\dpar\psi_n} + \dd\bar\psi_n\frac{\dpar}{\dpar\bar\psi_n}-\frac{(n-1)}{n}\,\frac{\dd\psi_n}{\psi_n}\ ,
\end{equation}
and the Riemann curvature is 
\begin{equation}\label{5.15}
\CR^{}_{\Gamma_n}=\nabla^{2}_{\Gamma_n}=\frac{2\pi(n-1)}{n}\,\delta (\rho)\,\dd\psi_n\wedge\dd\bar\psi_n\ ,
\end{equation}
where the delta-function $\delta (\rho)$ indicates the singularity of curvature at the point $\psi_n=0=z$.

Summing up, we obtain that the eigenfunctions $\Psi_+(n)$ of the Hamilton operator of quantum harmonic oscillator have the form \eqref{5.1} and define a fluctuating two-dimensional surface $\C/\Z_n\cong \R^2/\Z_n$ (standing wave) in the space $L_\C^+\cong\C^2$ with coordinate $\psi_n$ on $\C/\Z_n$, metric \eqref{5.12}, Levi-Civita connection \eqref{5.14} and curvature \eqref{5.15}. For $n=0$, the solution $\Psi_+(0)$ describes the rotation of the basis $v_+^c(t)$ in fibres of the bundle $L_\C^+\to\R^2$ and this rotation with a constant frequency does not depend on $n=0,1,...\ $. Using $\tau =\omega t$ we see that in one round around the circle in $\R^2$, the circle in $\R^2/\Z_n$ is walked $n$ times and the basis $v_+^c$ in fibres $\C$ rotates by $1/2$ of the circle, which in total gives the energy of the state $\Psi_+(n)$. The eigenfunction $\Psi_-(n)$ of the antiparticle has the same positive energy and opposite quantum numbers $\ql$ and $\qv$, both parametrized by the fundamental group $\pi_1(\R^2\setminus\{0\})=\pi_1(S^1) =\Z$.

\bigskip

\noindent 
{\bf\large Acknowledgments}

\noindent
I am grateful to Tatiana Ivanova for useful remarks.

\bigskip


\end{document}